\documentclass[
reprint,
showpacs,preprintnumbers,
amsmath,amssymb,
aip,apl,
]{revtex4-1}

\usepackage{amsthm}
\usepackage{color}
\usepackage{graphicx}
\usepackage{dcolumn}
\usepackage{bm}
\usepackage{upgreek}
\usepackage{txfonts}
\usepackage{tikz}
\usepackage{pgffor}
\usepackage{verbatim}
\usepackage{pstricks}
\usepackage[colorlinks, breaklinks=true, linkcolor=blue, citecolor=blue, linktocpage=true]{hyperref}
\newtheoremstyle{query}%
{}{}%space above/below
{\color{magenta}}%body style
{}%heading indent
{\sffamily\bfseries}{:}{12pt}%heading style/punctuation/space after
{}% head spec
\theoremstyle{query}

\makeatletter
\AtBeginDocument{\@ifpackageloaded{natbib}{\ifNAT@numbers\if@filesw\immediate\write\@auxout{\string\global\string\NAT@numberstrue}\fi\fi}{}
\let\($\let\)$}
\makeatother

\usepackage{amstext}

\begin{document}

\title{Thermoelectric refrigerator based on asymmetric surfaces of a magnetic topological insulator}

\date{\today}
\author{Takahiro Chiba}
 %\email{t.chiba@fukushima-nct.ac.jp}
 \affiliation{National Institute of Technology, Fukushima College, 30 Nagao, Kamiarakawa, Taira, Iwaki, Fukushima, 970-8034, Japan.}
%\author{Saburo Takahashi}
% \affiliation{Advanced Institute for Materials Research, Tohoku University, Sendai, Miyagi 980-8577, Japan}
\author{Takashi Komine}
 \affiliation{Graduate School of Science and Engineering, Ibaraki University, 4-12-1 Nakanarusawa, Hitachi, Ibaraki 316-8511, Japan}

\begin{abstract}
%------------------------------
Thermoelectric (TE) refrigeration such as Peltier cooler enables a unique opportunity in electric energy to directly convert thermal energy. Here, we propose a TE module with both refrigeration and power generation modes by utilizing asymmetric surfaces of a magnetic topological insulator (quantum anomalous Hall insulator) with a periodic array of hollows filled with two different dielectrics. Based on the Boltzmann transport theory, we show that its efficiency, i.e., the dimensionless figure of merit $ZT$ exceeds 1 in the low-temperature regime below 300~K. The proposed device could be utilized as a heat management device that requires precise temperature control in small-scale cooling.
%------------------------------
\end{abstract}

%\pacs{}

\maketitle

\section{Introduction}

Thermoelectric (TE) devices are used in a wide range of applications related to solid-state based power generation and refrigeration. In particular, the TE refrigeration such as Peltier cooler has drawn attention due to a CO$_2$--free cooling technology for automotive applications, computer processors, refrigeration of biological samples, and various heat management systems. \cite{DiSalvo99,Tritt11} 
The primary advantages of a Peltier cooler compared to a traditional vapor-compression refrigerator are flexibility and compactness owing to the lack of moving parts, enabling  applications for small-scale cooling. TE cooling technology is based on the Peltier effect in TE materials in which an electric current drives heat flow and creates the temperature difference at the hot and cold ends of a system. 

The efficiency of TE energy conversions is evaluated by the dimensionless figure of merit $ZT$. \cite{Goldsmid64,Tritt11}
Over the past several years, many new materials have been investigated for their use as TE materials with high $ZT$. \cite{Urban19}  So far, tetradymite--type chalcogenides such as Bi$_{2}$Te$_{3}$ have been well known as a good TE material with $ZT\approx1$, \cite{Ni05,Poudel08,Zahida10,Maassen13,Muchler13,Wickramaratne15} but have also drawn much attention as three-dimensional topological insulators (3D TIs) in recent years. \cite{Hasan10} 3D TI is an electronic bulk insulator but has a linear energy dispersion near a single band-touching (Dirac) point on the surface due to strong spin--orbit interaction. Recently, an ideal two-dimensional (2D) Dirac surface state in 3D TIs with a highly insulating bulk has been observed in ${\rm (Bi_{1-x}Sb_{x})_{2}Te_{3}}$ (BST) and ${\rm Bi_{2-x}Sb_{x}Te_{3-y}Se_{y}}$ (BSTS). \cite{Ando13} By focusing on the TI surface states, some potential systems and devices to realize high-performance thermoelectrics so far have been theoretically proposed.  \cite{Ghaemi10,Tretiakov11,Takahashi12,Xu14,Osterhage14,Gooth15,Shi15,Chiba19JAP} 

According to the previous studies \cite{Tretiakov11,Takahashi12,Chiba19JAP}, one of the simplest approaches to achieve a high $ZT$ is the introduction of an surface band gap on the TI surface. \cite{Tretiakov11,Takahashi12}
A system with massive Dirac electrons on a gap-opened TI surface can be realized by  hybridization of the top and  bottom surfaces. \cite{Souma12,Neupane14} This mechanism is applied to 3D TIs with many holes in the bulk \cite{Tretiakov11} or to a superlattice made from a 3D TI and an empty layer. \cite{Fan12} 
A recent experiment has observed a large Seebeck coefficient in a ultrathin film of BSTS owing to the surface gap-opening by the hybridization effect. \cite{Matsushita17} In contrast, since a surface band gap is also induced by a magnetic perturbation that breaks the time-reversal symmetry, the application of a magnetic field should be the simplest approach. However, magnetic fields of \(
{\sim}10\)~T induce a very small subgap (of the order of several meV) in the surface of 3D TIs. \cite{Analytis10} An alternative approach is magnetic doping into a 3D TI \cite{Checkelsky12,Lee15} or making ferromagnet contact with magnetic proximity effect, \cite{Jiang15,Hirahara17,Chiba17,Mogi19} which can induce a large surface band gap of the order of 100~meV. It is known that ferromagnetism in the magnetically doped 3D TIs can be developed through  the carrier-mediated Ruderman--Kittel--Kasuya--Yosida (RKKY) mechanism and/or the carrier-independent bulk Van Vleck mechanism. \cite{Kou15,Tokura19} In particular, the gap-opened magnetic TI surface exhibits the quantum anomalous Hall effect, characterizing the topological nature of 2D massive Dirac electrons, \cite{Chang13} and thus would be expected as a new platform for studying magneto--thermoelectric properties. 

In this paper, we propose a TE module utilizing asymmetric surfaces of a magnetic TI  (quantum anomalous Hall insulator) in which a periodic array of hollows filled with two different dielectrics is introduced. A pair of these two surfaces that are adjoined with each other acts as a $\Pi$-shaped $p$-$n$ junction with ambipolar conduction \cite{Chen12,Kim14}, which can be regarded as a thermocouple consisting of two dissimilar TE materials \cite{Chiba19}. Thus, a serial connection of the thermocouple operates as a TE module with both refrigeration and power generation modes. By using the Boltzmann transport theory at finite temperatures, we show that $ZT$ exceeds 1 in the low-temperature regime below 300~K. The proposed device could be utilized as a heat management device that requires precise temperature management. 

\section{Device proposal}

\begin{figure}[ptb]
\begin{centering}
\includegraphics[width=0.4\textwidth,angle=0]{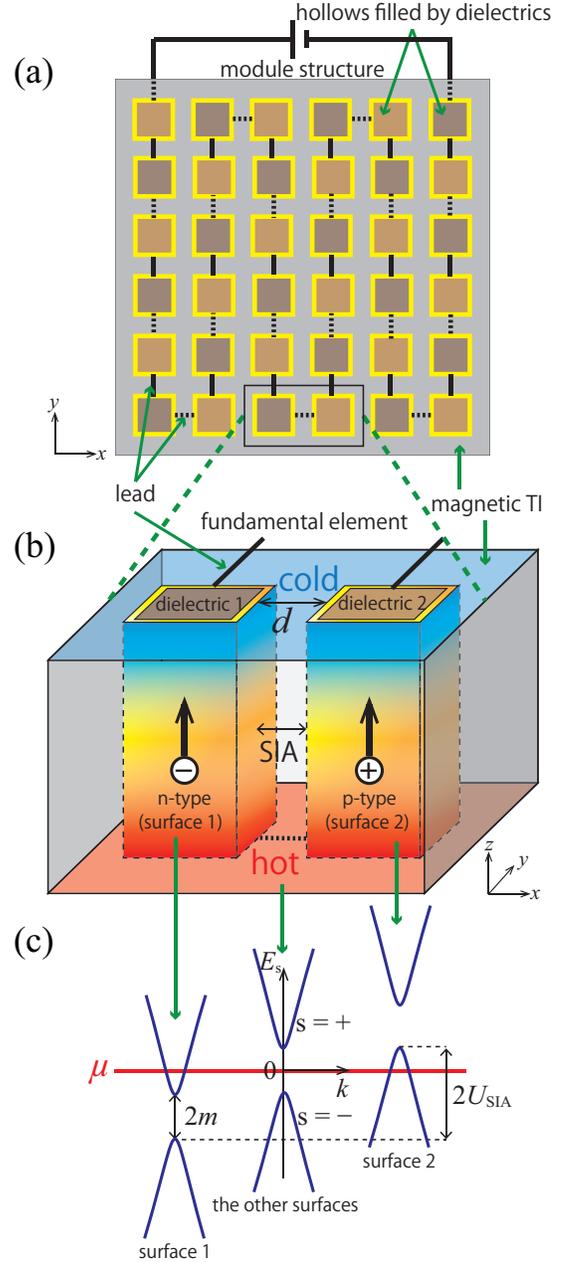} 
\par\end{centering}
\caption{
(a) Schematic illustration (top view) of the TE module made of a film of magnetic TI. A periodic array of small square hollows filled with two different dielectrics is introduced into the magnetic TI. Each hollow harbors  gap-opened metallic surface states (yellow lines) and is connected in series by leads (black solid and dashed lines). Refrigeration mode is shown here.
(b) Schematic geometry of the fundamental element (topological thermocouple) 
consisting of two connected hollows with different dielectrics (1 and 2), possessing the $p$- and $n$-types metallic surface states. $d$ is the distance between the two adjoined surfaces.
(c) Corresponding $k$-dependent surface band dispersions around the $\Gamma$ point are depicted by blue lines in which $\mu$ denotes the chemical potential at equilibrium and $U_{\rm SIA}$ describes the structure inversion asymmetry (SIA) between the adjoined two surfaces due to band bending induced by the dielectrics. 
}
\label{fig:device}
\end{figure}

Here, we designs a TE module utilizing asymmetric surfaces of a magnetic TI. In Fig.~\ref{fig:device}, we summarize the concept of the proposed device. Figure~\ref{fig:device}~(a) shows the TE module made of a film of magnetic TI (quantum anomalous Hall insulator \cite{Chang13}) in which a periodic array of hollows filled with two different dielectrics is introduced. Such dielectric-filled hollows give rise to gap-opened metallic surface states, as shown in Fig.~\ref{fig:device}~(c) by yellow lines. In this paper, we call a pair of the two hollows connected by a lead a ``topological thermocouple,'' and its structure is schematically illustrated in Fig.~\ref{fig:device}~(b).  
A pair of these two surfaces that are adjoined with each other acts as a $\Pi$-shaped $p$-$n$ junction with ambipolar conduction, which can be regarded as a thermocouple consisting of two dissimilar TE materials. 
It is worth noting that recent experiments demonstrated one surface with positive carriers and the opposite surface with negative carriers in a heterostructure based on a magnetically doped 3D TI.\cite{Fan16} The difference in carrier types originates from the structure inversion asymmetry (SIA) between the two adjoined surfaces in Fig.~\ref{fig:device}~(b), which is induced by the band bending imposed by the dielectrics. \cite{Wang15,Chiba19} 
The effective Hamiltonian for a pair of  adjoined surfaces is 
\begin{align}
\mathcal{H}_{\mp}(k)=\mp\hbar v_{\rm F}\left(  \sigma_xk_y - \sigma_yk_x\right) + m\sigma_z \mp U_{\rm SIA}\sigma_0
,\label{H2d2}
\end{align}
where $\mp$ indicates TI surfaces attached to dielectric 1 ($-$) and 2 ($+$), $U_{\rm SIA}$ denotes the SIA between the two adjoined  surfaces, $\sigma_0$ is the identity matrix, and $m$ corresponds to the surface band gap. For simplicity, we do not consider the particle--hole asymmetry in the surface bands and assume that the gap-opened surface states have symmetric energy dispersions: $E_{s}^{\pm}(k) = \mp s\sqrt{(\hbar v_{\rm F}k)^2 + m^2} \mp U_{\rm SIA}$ in which $s=\pm$ labels the upper/lower surface bands, which are schematically depicted in Fig.~\ref{fig:device}~(c). 
Thus, a serial connection of the topological thermocouple can operate as a TE module with both refrigeration and power generation modes. 
To fabricate the proposed device, we might utilize the nanoimprint lithography which enables us to create a mold for making convex hollows. If the thickness is about 10~$\mu$m, many submicron hollows can be made by the mold. After molding, the electrode pattern is formed by photolithography in the submicron-scale.

%=================<fig1>===================== 
%============================================

\section{Thermoelectric properties}

To model the TE properties of the proposed device, we assume the emergence of ionic defects in the bulk of the TI as well as on its surface, taking into account the effect of element substitution of the 3D TI for systematic control of the Fermi levels \cite{Shi15,Chiba19JAP}. Accordingly, based on the Boltzmann transport theory, we treat ionic disorder as a randomly distributed Coulomb-type long-range potential: $V_{\rm c}(\mathbf{r})=(e^2/\epsilon)\sum_{i}1/|\mathbf{r} - \mathbf{R}_i|$ with impurity concentration $n_{\rm c}$ and the effective lattice dielectric constant $\epsilon$. \cite{Chiba19JAP} 
Assuming an ensemble averaging over random uncorrelated impurities up to lowest order in the scattering potential $V_{\rm c}(r)$, we can obtain the transport relaxation time \cite{Chiba19JAP}
\begin{align}
\tau(E_{s}^{\pm}) =\tau_{\mathrm{c}}^{(0)}(E_{s}^{\pm})\left[1+3\frac{m^2}{(E_{s}^{\pm})^2}\right]^{-1}
,\label{tau c2}
\end{align}
where $\tau_{\mathrm{c}}^{(0)}(E_{s}^{\pm})=E_{s}^{\pm}/(\pi^2\hbar v_{\rm F}^2n_{\rm c})$ denotes the transport relaxation time for the gapless surface state. 

According to the linear response theory, charge (${\bf j}_c^p$) and thermal (${\bf j}_Q^p$) currents ($p=-$ for electron and $p=+$ for hole) can be described by linear combinations of an electric field ${\bf E}$ and a temperature gradient $\bm{\nabla}T$:
\begin{align}
\begin{pmatrix} {\bf j}_c^p \\[3pt] {\bf j}_Q^p
\end{pmatrix}
=
\sigma_p
\begin{pmatrix} 1 & S_pT \\[3pt] \varPi_p & \kappa_pT/\sigma_p% + \varPi_p^2
\end{pmatrix}
\begin{pmatrix} {\bf E} \\[3pt] -\bm{\nabla}T/T
\end{pmatrix}
,\label{TE coefficient p}
\end{align}
where the electrical sheet conductance $\sigma_p=e^2L_0^p$ (in units of S = $\Omega^{-1}$) with electron charge $-e~(e>0)$, the Seebeck coefficient $S_p=L_1^p/(eL_0^pT)$ (in units of V\,K$^{-1}$), the Peltier coefficient $\varPi_p=S_pT$ (in units of V), and the thermal sheet conductance $\kappa_p=[L_0^pL_2^p-(L_1^p)^2]/(L_0^pT)$ (in units of W\,K$^{-1}$). For the application of ${\bf E}$ and $\bm{\nabla}T$ along the $x$ direction, the coefficients $L_n^p$~$(n=1,2,3)$ are obtained by
\begin{align}
L_n^p={}&\sum_{s}\int \frac{d{\bf k}}{(2\pi)^2}\tau(E_{s}^{\pm})(  {\bf v}_{s}^{\pm})_{x}^2\left(-\frac{\partial f^{(0)}}{\partial E_{s}^{\pm}}\right)p^n(  \mu-E_{s}^{\pm})^n,\label{Ln}
\end{align}
$\mathbf{v}_{s}^{\pm}=\bm{\nabla}_{\mathbf{k}}E_{s}^{\pm}/\hbar$ is the group velocity of carriers, $f^{(0)}$ the equilibrium Fermi-Dirac distribution, and $\mu$ the chemical potential measured from the Dirac point ($E_{s}^{\pm}=0$) of the original gapless surface band.
Due to the heat transport by phonons, we need to include the thermal conductivity of phonons $\kappa_{\rm ph}$ (in units of W\,K$^{-1}$\,m$^{-1}$) in the definition of $ZT$. \cite{Goldsmid64} 
In the proposed device, the surface band structures of two adjoined surfaces are assumed to be symmetric so that $ZT$ is equivalent to that of the individual surfaces and becomes a maxim. By using Eq.(\ref{Ln}), the figure of merit on the TI surfaces is therefore given by \cite{Chiba19JAP}
\begin{align}
ZT = \frac{\sigma_p S_p^2T}{\kappa_p+d\kappa_{\rm ph}} = \frac{(L_1^p)^2}{L_0^p(L_2^p+d\kappa_{\rm ph}T)-(L_1^p)^2}
,\label{ZT2}
\end{align}
where $d$ is the distance between the two adjoined surfaces, taking the role of a factor related to the surface-to-bulk ratio. 

%=================<fig2>=====================
\begin{figure}[ptb]
\begin{centering}
\includegraphics[width=0.45\textwidth,angle=0]{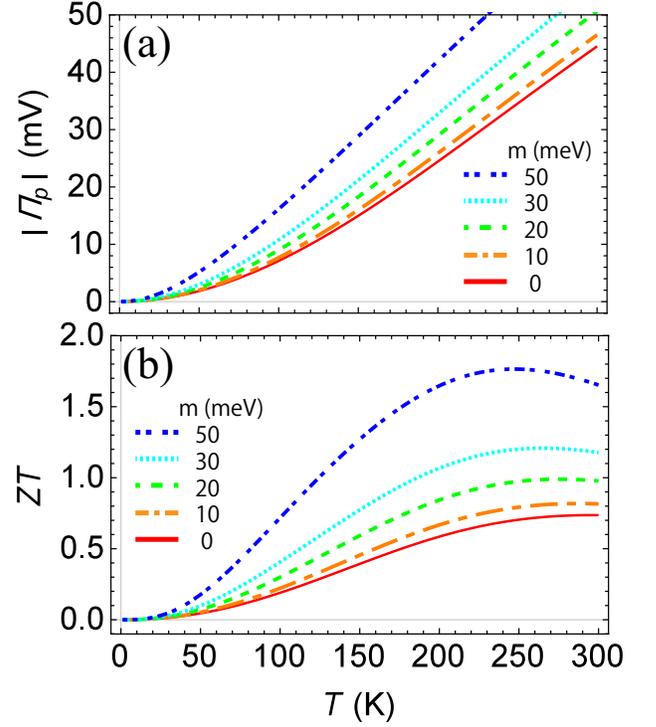} 
\par\end{centering}
\caption{(a) Peltier coefficient and (b) thermoelectric figure of merit arising from a screened Coulomb impurity as a function of $T$ for different $m$. In this plot, we set $\mu=65$~meV and $n_{\rm c}=10^{10}$~cm$^{-2}$. 
The details of the calculations are given in the text.
}
\label{fig:TE coefficient c}
\end{figure}
%============================================

Figure~\ref{fig:TE coefficient c}~(a) shows the calculated Peltier coefficient $|\Pi_p|$ as a function of $T$ for different values of $m$. As seen, the Peltier coefficient increases with increasing both $T$ and $m$. In this plot, based on the experiment in Ref.~\onlinecite{Fan16}, we assume a carrier density $5.0\times10^{11}$~cm$^{-2}$, which corresponds to $\mu\approx65$~meV, and take $v_{\rm F}=4.0\times10^{5}$~m\,s$^{-1}$ as reported in Ref.~\onlinecite{Arakane12}. To decrease the heat transport due to phonons, we assume a thin film of 3D TI of thickness $d = 10$~nm. It is noting that the topological surface dominates transport in thin films of a 3D TI with $d\leq14$~nm was reported in recent experiments. \cite{Matsushita19} Figure~\ref{fig:TE coefficient c}~(b) shows the calculated thermoelectric figure of merit $ZT$ as a function of $T$ for different values of $m$. In contrast to the Peltier coefficient, $ZT$ has a peak in the temperature range from 200 to 300~K. This is understandable because when the surface band gap opens, the thermal currents driven by the Peltier effect and a thermal gradient partially cancel through the relation~(\ref{TE coefficient p}) for ${\bf E}={\bf 0}$: ${\bf j}_Q = \left( L_2^p - \sigma_p\varPi_p^2 \right)\left(  -\bm{\nabla}T/T\right)$, leading to the maximization of $ZT$. Since the proposed device enhances the $ZT$ in small scales in terms of $d$, we suggest that our TE module could be combined with optoelectronic devices such as cooling laser diodes that require precise temperature changes \cite{DiSalvo99} as well as be utilized for refrigeration of biological samples that require sensitive temperature control at localized spots.

\section{Summary}

In summary, we have proposed a TE module with both refrigeration and power generation modes by utilizing asymmetric surfaces of a magnetic topological insulator (quantum anomalous Hall insulator). A pair of these two surfaces that are adjoined with each other acts as a $\Pi$-shaped $p$-$n$ junction with ambipolar conduction, which can be regarded as a thermocouple consisting of two dissimilar TE materials. Thus, a serial connection of the thermocouple operates as a TE module. By using the Boltzmann transport theory, we demonstrated that its efficiency, i.e., $ZT$ exceeded 1 in the low-temperature regime below 300~K. The proposed device could be utilized as a heat management device that requires sensitive temperature changes in a wide variety of applications for small-scale cooling.

\section*{Acknowledgments}

The authors thank S. Takahashi, S. Y. MatsushitaK. Tanigaki, and Y. P. Chen for valuable discussions. This work was supported by Grants-in-Aid for Scientific Research (Grant No.~20K15163  and No.~20H02196) from the JSPS.

\section*{Data availability}

The data that support the findings of this study are available from the corresponding author upon reasonable request.

%\appendix*
%
%\renewcommand{\thefigure}{\hbAppendixPrefix\arabic{figure}}
%\setcounter{figure}{0}
%\renewcommand{\thetable}{\hbAppendixPrefix\arabic{table}} 
%\setcounter{table}{0}

%\clearpage

\end{document}